\documentclass[english]{article}
\usepackage{graphicx}
\usepackage[margin=1.35in]{geometry}
\usepackage{algorithm, algorithmic}

\usepackage{amsmath,amsthm,amssymb}
\usepackage{mathtools}
\usepackage{tikz}
\usetikzlibrary{matrix,arrows,cd,calc}
\usepackage{mathbbol, graphicx, float}
\usepackage{verbatim}
\usepackage{setspace}
\usepackage{subcaption}

\usepackage{hyperref}
	\hypersetup{colorlinks=true, citecolor=blue, linkcolor=red, urlcolor=pink, pdfstartview={FitH}}

\usepackage[numbers,sort&compress]{natbib}

\newtheorem{theorem}{Theorem}
 
\newtheorem{lemma}{Lemma}

\newtheorem{corollary}{Corollary}
\newtheorem{definition}{Definition}

\newcommand\term{\emph}
\newcommand\module{\mathbf}
\let\subset=\subseteq

\newcommand{\out}{\mathrm{out}}

\newcommand{\LocateAndFree}{\operatorname{\text{\textsc{LocateAndFree}}}}

\title{A Universal In-Place Reconfiguration Algorithm for Sliding Cube-Shaped Robots in a Quadratic Number of Moves\footnote{This paper extends and subsumes an earlier preprint by Abel and Kominers~\cite{abel2008universal}.}}

\author{
Zachary Abel\thanks{%
MIT, \texttt{zabel@mit.edu}}
\and
Hugo A. Akitaya\thanks{%
U.~Mass.~Lowell, \texttt{hugo\_akitaya@uml.edu}}
\and
Scott Duke Kominers\thanks{%
Harvard University and a16z crypto, \texttt{kominers@fas.harvard.edu}}
\and
Matias Korman\thanks{%
Siemens Electronic Design Automation, \texttt{matias.korman@siemens.com}}
\and
Frederick Stock\thanks{%
U.~Mass.~Lowell, \texttt{frederick\_stock@student.uml.edu}} 
}

\date{March 2024}

\begin{document}

\maketitle
\begin{abstract}
In the modular robot reconfiguration problem, we are given $n$ cube-shaped \textit{modules} (or \textit{robots}) as well as two \textit{configurations}, i.e., placements of the $n$ modules so that their union is face-connected. The goal is to find a sequence of moves that reconfigures the modules from one configuration to the other using ``sliding moves,'' in which a module slides over the face or edge of a neighboring module, maintaining connectivity of the configuration at all times.

For many years it has been known that certain module configurations in this model require at least $\Omega(n^2)$ moves to reconfigure between them. In this paper, we introduce the first universal reconfiguration algorithm---i.e., we show that any $n$-module configuration can reconfigure itself into any specified $n$-module configuration using just sliding moves. Our algorithm achieves reconfiguration in $O(n^2)$ moves, making it asymptotically tight. 
We also present a variation that reconfigures \emph{in-place}, it ensures that throughout the reconfiguration process, all modules, except for one, will be contained in the union of the bounding boxes of the start and end configuration.
\end{abstract}

\section{Introduction}\label{sec:intro}
A \emph{modular self-reconfigurable robotic system} is a set of robotic units (called \emph{modules}) that can communicate, attach, detach and move relative to each other; various models of such robots have received considerable attention by the computational geometry community~\cite{abel2008universal,aloupis2009linear,aloupis2008reconfiguration,dumitrescu2004motion,DP,akitaya2021universal, sung2015reconfiguration}. When analyzing a model of modular robots, the typical goal is to find a \emph{universal reconfiguration algorithm} defined as follows: 
A \emph{(reconfigurable robot) configuration} is an arrangement of modules in space that is required to be connected. A \emph{move} is a local rearrangement involving one module that transforms one configuration into another.
During a move, the set of stationary modules is also required to be connected (known as the \emph{single backbone condition}~\cite{dumitrescu2004motion}).

We say that reconfiguration is \emph{universal} if given any two configurations $s$ and $t$, there is always a series of moves that reconfigures $s$ to $t$; an algorithm that computes this sequence of moves is a \emph{universal reconfiguration algorithm}. For many models of reconfigurable robots, universal reconfiguration is impossible \cite{akitaya2021universal,sung2015reconfiguration}, and furthermore, it is often NP or PSPACE-complete to determine if one can even reconfigure between two configurations~\cite{a.akitaya_et_al:LIPIcs.SoCG.2021.10}. 

\begin{figure}[ht]
    \centering
    \begin{subfigure}[t]{.45\textwidth}
        \centering
        \includegraphics[scale=.7]{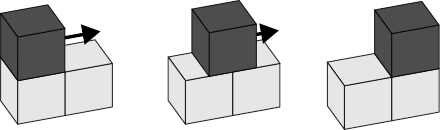}
        \caption{Slide Move}
    \end{subfigure}\hfill
    \begin{subfigure}[t]{.45\textwidth}
        \centering
        \includegraphics[scale=.7]{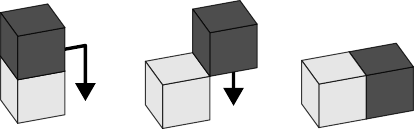}
        \caption{Rotation Move}
    \end{subfigure}
    \caption{Two types of allowed moves in the sliding cube model}
    \label{fig:Cubemoves}
\end{figure}

We consider the \emph{sliding (hyper-)cube model}, in which each module is a (hyper-)cube, and a configuration comprises a placement of the cubes into lattice-aligned positions so that the interior of their union is connected.
Two ($d$-dimensional hyper-)cubes are \emph{adjacent} if they share a face (i.e., a $(d-1)$-dimensional facet). 
A module can slide along a face of an adjacent module, either moving to be adjacent to a new module or rotating around a corner of a module (see Figure~\ref{fig:Cubemoves}). 
The \emph{free-space requirement} for a move is the set of lattice positions that are required to be empty for a move to be collision-free.

This sliding cube model has attractive properties relative to other popular reconfigurable robot systems. In the \textit{pivoting model}, robots rotate around a shared edge instead of sliding on a face~\cite{sung2015reconfiguration}; this model requires more free space, making reconfiguration more difficult or only possible in limited cases~\cite{akitaya2021universal}. In the \emph{crystalline model}~\cite{aloupis2009linear}, robots move via expansions and contractions, and universal reconfiguration requires $2\times 2\times 2$ \emph{meta-modules}, small sub-arrangements of modules which are treated as the atomic units.
Furthermore, $2\times 2\times 2$ crystalline meta-modules can simulate sliding moves and, thus, any reconfiguration algorithm for the sliding cube model can also be used for the crystalline model~\cite{aloupis2009realistic}.\footnote{The analogous statement is true for higher-dimensional systems.}

The problem of sliding cube reconfiguration is fairly well understood in two dimensions: Dumitrescu and Pach~\cite{DP} were the first to show a universal reconfiguration algorithm for $n$ modular sliding $2$-cubes, using $O(n^2)$ moves.
Moreno and Sacrist\'an~\cite{moreno2020reconfiguring} adapted Dumitrescu and Pach's algorithm to run \emph{in-place}, i.e., at any time only one robot is outside the union of the bounding boxes of the start and target configurations.
Recently, Akitaya et al.~\cite{a.akitaya_et_al:LIPIcs.SWAT.2022.4} showed that finding the shortest reconfiguration (in terms of number of moves) is NP-hard in 2D. They also improved on Moreno and Sacrist\'an's algorithm by maintaining in-place property while simultaneously making the algorithm input-sensitive. Effectively, they reduced the number of moves to $O(\overline{P}n)$ moves, where $\overline{P}$ is the maximum perimeter of the two bounding boxes.

For three (and higher) dimensions, sliding cube configuration has also received significant attention, yet not so much is known. The sliding cube model was first described by Fitch, Butler, and Rus~\cite{fitch2003reconfiguration}. They proposed a simple universal reconfiguration strategy for a configuration $C$ of $n$ modules: find a module $\module m$ on the outer boundary of $C$ that will not break connectivity if removed, move $\module m$ on top of a fixed extreme position on the boundary, and repeat this operation $n-2$ more times. The Fitch, Butler, and Rus approach would lead to a reconfiguration algorithm that requires $O(n^2)$ moves. It is known that $\Omega (n^2)$ moves are sometimes necessary for reconfiguration in 3D (explicit construction shown in~\cite{miltzow2020hiding}), which would make the Fitch, Butler, and Rus algorithm optimal.

Critically, Fitch, Butler, and Rus~\cite{fitch2003reconfiguration} rely on the continuous existence of the desired module $\module m$. However, Miltzow et al.~\cite{miltzow2020hiding} recently presented a configuration where no such module exists. In other words, there are configurations of sliding cubes where no module on the outer boundary of $C$ can move without breaking connectivity---implying that a more complex approach is necessary. 

In 2008, Abel and Kominers announced a universal reconfiguration algorithm for dimensions $3$ and higher in an arXiv 
preprint~\cite{abel2008universal}. In essence, their algorithm requires $O(n^2)$ moves to adaptively ensure that the condition of Fitch, Butler, and Rus~\cite{fitch2003reconfiguration} is satisfied, which leads to an overall $O(n^3)$ algorithm. Their result was never formally published and unfortunately, the analysis has some minor flaws.\footnote{Historical note: This document subsumes and further extends the results presented in the Abel--Kominers preprint~\cite{abel2008universal}. We acknowledge that a lot of time has passed since \cite{abel2008universal} was initially posted online, so in the remainder of this document, we treat it as a separate document.} Parallel to our work, another group of researchers announced a universal in-place reconfiguration algorithm for three and higher dimensions~\cite{inPlaceCompact23}. Their algorithm is input-sensitive, and the number of moves is bounded by the overall sum of coordinates of modules in both configurations (values that range from $\Omega(n^{4/3})$ to $O(n^{2})$).

\paragraph*{Contribution} 
After we present the key definitions and model framework in Section~\ref{sec:def}, Section~\ref{sec:framework} introduces topological properties that are foundations of the algorithms presented in this paper. In Section~\ref{sec:algorithm}, we show that a slight modification of the Abel--Kominers~\cite{abel2008universal} algorithm achieves universal reconfiguration for the sliding (hyper-)cubes model. (In Section~\ref{sec_error} we identify and fix a small mistake in the Abel--Kominers~\cite{abel2008universal} manuscript.) Moreover, by improving the analysis (and with our minor corrections to the algorithm), we can reduce the required moves to $O(n^2)$, making the algorithm optimal. 

In Section~\ref{sec:inPlace}, we modify the algorithm further to obtain a new in-place algorithm for sliding cube reconfiguration; by \emph{in-place} we mean during reconfiguration, moving modules stay within $O(1)$ distance of the union of the bounding boxes of the start and target configurations, as in~\cite{moreno2020reconfiguring}. In addition, our algorithm is input-sensitive, similar to Akitaya et al.'s result \cite{a.akitaya_et_al:LIPIcs.SWAT.2022.4}. A natural $3$-dimensional extension of their $O(\overline{P}n)$ algorithm would use $O(Vn)$ moves ($V$ being the volume of the configuration) but our algorithm requires fewer---the bound is closer to $O(\overline{P}n)$ than $O(Vn)$; a formal statement is in Section~\ref{sec:inPlace}. 

\section{Preliminaries: The Sliding Cube Model}\label{sec:def}
Let $M$ be a set of $n$ distinct $d$-dimensional hypercube modules.
A \emph{labeled configuration} of $M$ is an injective function from $M$ to the set of unit cells in the $d$-dimensional hyper-cube lattice. (Here, by a \textit{cell}, we mean an axis-aligned unit cube with vertices on the integer grid.)
The image of a labeled configuration in the hypercubic lattice is called an \emph{unlabeled configuration}.

For an unlabeled configuration $C$, we say that a lattice cell is \emph{occupied} if it is in $C$; the cell is \emph{empty} otherwise. We sometimes abuse terminology by referring to the occupied cells of $C$ as simply the ``cells of $C$.''
We refer to the $(d-1)$- and $(d-2)$-dimensional facets a lattice cell as \emph{faces} and \emph{edges}, respectively.
Two cells are \emph{adjacent} if they share a face.

Let $G_C$ be the graph whose vertices are the cells of $C$ with edges connecting pairs of cells that share one face.
We say that $C$ is connected if $G_C$ is connected. 
Note that a connected configuration is a polyhypercube (a generalization of a polyomino to dimension $d$).
Although we focus on unlabeled configurations, for clarity we may refer to them by way of their labeled counterparts, using language such as ``move module $\module m$ from cell $a$ to cell $b$.'' 
We also sometimes abuse notation by referring to a cell by the module that occupies it in a configuration, for example, saying that two modules are ``adjacent'' in a configuration if the cells they occupy are adjacent, or using the notation ``$C\setminus\{\module m\}$'' to denote the configuration obtained by subtracting from $C$ the cell occupied by a module $\module m$. 

Note that the complement $\overline{C}$ of $C$ might be disconnected, and $\overline{C}$ has exactly one unbounded component.
Let $\partial C$ denote the boundary of $C$ and let the \emph{outer boundary} of $C$ be the boundary of the unbounded component of $\overline{C}$.
Let $B_\out(C)$ be the set of modules in $C$ that have at least one face on the outer boundary.
The faces of $C$ in $\partial C$ comprise the \emph{boundary faces} of $C$. We define the outer boundary of components of $\overline{C}$ in a symmetric fashion; therefore, the boundary faces of the unbounded component of $\overline{C}$ are the same as the boundary faces of $C$. 

A \emph{move} transforms a configuration $C$ into a configuration $C'$ that differs from $C$ by the position of a single module $\module m$; we refer to $\module m$ as the \emph{moving} module and say that the other modules are \emph{stationary}. Only certain types of moves are considered valid, as we specify next.
If for any starting configuration $C$ and target configuration $C'$ there is a sequence of valid moves that can reconfigure $C$ into $C'$, then we say that the model is \textbf{universal}; an algorithm that performs such reconfiguration on arbitrary input is likewise called \emph{universal}. 

We say that $\module m$ is \emph{articulate} in configuration $C$ if the cell occupied by $\module m$ is a cut vertex of $G_C$; $\module m$ is \emph{nonarticulate} otherwise.
We require the \emph{single backbone condition}: for any configuration $C$, all valid moves must result in a configuration $C'$ with $C\cap C'$ connected; that is, any valid move must leave the configuration of cells occupied by stationary modules connected.
The single backbone condition is equivalent to requiring that only nonarticulate modules move.

In the \textit{sliding model}, two types of moves are allowed: slides and rotations. These \emph{sliding moves} are as follows (refer to Figure~\ref{fig:Cubemoves}, where the moving module $\module m$ is shown in dark grey):
\begin{itemize}
	\item A \emph{slide} moves $\module m$ from a cell $a$ to an adjacent empty cell $b$, and requires that there are adjacent occupied cells $a'$ and $b'$ such that $a$ is adjacent to $a'$ and $b$ is adjacent to $b'$.
  \item A \emph{rotation} moves $\module m$ from a cell $a$ to an empty cell $b$ where $a$ and $b$ share a common edge $e$, and are both adjacent to an occupied cell $c$, and requires that the cell $d\notin\{a,b,c\}$ that contains $e$ is empty. Note that every edge $e$ is incident to exactly $4$ cells.   
\end{itemize}

\begin{figure}
    \centering
    \begin{subfigure}[t]{.22\textwidth}
        \centering
        \includegraphics[scale=.75]{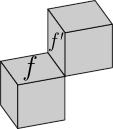}
        \caption{Slide-Adjacent}
    \end{subfigure}\hfill
    \begin{subfigure}[t]{.22\textwidth}
        \centering
        \includegraphics[scale=.75]{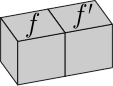}
        \caption{Slide-Adjacent}
    \end{subfigure}\hfill
    \begin{subfigure}[t]{.40\textwidth}
        \centering
        \includegraphics[scale=.75]{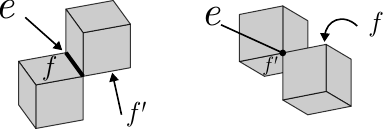}
        \caption{Not Slide-Adjacent}
    \end{subfigure}
    \caption{Examples of slide adjacency. In (a) and (b), $f$ and $f'$ are slide-adjacent; (c) shows how two faces $f$ and $f'$ can share an edge but not be slide-adjacent.}
    \label{fig:slideAdj}
\end{figure}

We now define a relationship called \emph{slide-adjacency} on the boundary faces of $C$ based on sliding moves.
Intuitively, this will allow us to argue about the possible positions that a module $\module m$ can occupy after performing a sequence of sliding moves on $C \setminus \module m$. We say that a nonarticulate module $\module m$ is \emph{attached} at a face $f$ if $f$ is in the common boundary between $\module m$ and $C\setminus\{\module m\}$.
We define a pair of boundary faces $f$ and $f'$ to be \emph{slide-adjacent} if they share an edge and, either (i) both are incident to an empty cell $a$ (Figure~\ref{fig:slideAdj}(a)) or (ii) $f$ is incident to an empty cell $a$ and $f'$ is incident to an empty cell $b$, and if a module attaches to $f$, it can move to $b$ with a single sliding move (Figure~\ref{fig:slideAdj}(b)). Two boundary faces $f$ and $f'$ could share an edge and yet a single sliding move cannot bring a module attached to $f$ to a position where it would be attached to $f'$.
This happens when an edge $e$ is \emph{pinched}, i.e., exactly two of the four cells containing $e$ are occupied and not adjacent (Figure~\ref{fig:slideAdj}(c)).

The \emph{slide-adjacency graph} of $C$ is the graph representing the slide-adjacency relations on the boundary faces of $C$.
Note that by definition, for every boundary face $f$ and edge $e$ contained in $f$ there is a unique boundary face $f'$ that is slide-adjacent to $f$ and shares $e$.

\section{Reconfiguration Framework}\label{sec:framework}

In Section~\ref{sec:algorithm} we revisit the $O(n^3)$-move universal reconfiguration algorithm of Abel and Kominers~\cite{abel2008universal} and modify it, showing that it actually performs $O(n^2)$ moves. Thus, we prove the following main theorem:

\begin{theorem}\label{thm:main}
  Given any two connected unlabeled configurations $C$ and $C'$ each having $n\ge 2$ modules, there exists a reconfiguration of $C$ into $C'$ using $O(n^2)$ sliding moves.
\end{theorem}

Similar to the approach of Dimitrescu and Pach~\cite{DP}, we prove our main result by showing that any
configuration can be reconfigured into a straight chain of modules, called the \emph{canonical configuration}. This suffices to prove the result, as it follows that any configuration $C$ can be reconfigured into this canonical straight
position, and may then be reconfigured into any other configuration $C'$. Note that the straight configuration may easily be relocated and reoriented in space by rotations and slides. Indeed, in Section~\ref{sec:inPlace}, we modify the reconfiguration algorithm to be in-place by placing the modules in a more compact form. However, before we can prove Theorem~\ref{thm:main} in Section~\ref{sec:algorithm}, we need several additional definitions and lemmata presented here.

\subsection{Structural Properties}\label{sec:Structural_Properties}
In this section, we show some structural properties that are the basis for our algorithms. For ease of description, the images focus on the case where $d=3$, but we note that our results are topological and thus extend to higher dimensions. Our analysis ignores the dependency of the dimension in the number of moves (alternatively, it considers that the dimension $d$ is constant). 

We also note that the properties listed below assume neither the start nor the end configuration fit in a $2$-dimensional plane. Thus, we assume neither configuration fits in a subspace of smaller dimension. If only one configuration fits in a $2$-dimensional subspace we virtually perform a single rotation to move a module outside the $2$-dimensional space. We use the modified configuration as the initial/target for our algorithm and then undo the move as the first or last step. If both configurations lie in the 2D subspace, we can use a known two-dimensional reconfiguration algorithm (such as~\cite{a.akitaya_et_al:LIPIcs.SWAT.2022.4}).

It is known that any connected graph $G$ on $n\ge 2$ vertices contains at least two distinct non-cut vertices; 
it follows immediately that any connected configuration $C$ on $n\ge 2$ modules contains at least two non-articulate modules.

\begin{figure}[ht]
  \centering
 \includegraphics[scale=1]{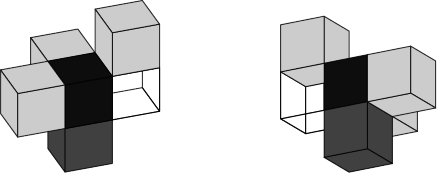}

  \caption{Lemma~\ref{lem:structure}: removing $\module x$ (black) disconnects
    $\module y$ (dark gray) from the boundary $B_\out(C)$. (Two views are
    presented.)}
  \label{fig:structure}
\end{figure}

\begin{lemma}\label{lem:structure}
  Suppose $\module x\in B_\out(C)$ is an articulate module, and further suppose that
  $\module x$ is adjacent to a module $\module y$ (along face $f$ of
  $\module x$) such that the connected component of
  $C\setminus\{\module x\}$ containing $\module y$ is disjoint from
  $B_\out(C)$. Without loss of generality, $\module y$ is the bottom neighbor of $\module x$.
  Then, as pictured in Figure~\ref{fig:structure}:
  \begin{enumerate}
    \item \label{it:i} The face $f^{\text{op}}$ of $\module x$ opposite $f$ is on the
        outer boundary of $C$;
    \item \label{it:ii} any module $\module w\neq \module y$ adjacent to $\module x$
        is in a component of $C\setminus\{\module x\}$ not disjoint from
        $B_\out(C)$; and
    \item \label{it:iii} $\module x$ is adjacent to at least one such module $\module
        w\neq \module y$. 
    \item \label{it:iv} Moreover, if a cell horizontally adjacent to $\module x$ is empty, then the cell directly above it is occupied by a module in $B_\out(C)$.
  \end{enumerate}
\end{lemma}

\begin{proof}
  Suppose part (\ref{it:i}) is false, meaning $f^{\text{op}}$ is not on the outer boundary of $C$.  Module $\module x$ has some face $g$ on the outer boundary, which is neither $f$ nor $f^{\text{op}}$.  Such a $g$ is edge-adjacent to $f$. Let $p$ be the empty cell adjacent to $\module x$ at $g$, and let $q$ be the cell not containing $\module x$ adjacent to both $p$ and $\module y$. Since $g$ is on the outer boundary, $p$ is empty. But since $\module y$ is not in $B_\out(C)$, $q$ must contain a module $\module{m}_q$. However, this means $\module y$ is adjacent to $\module{m}_q$, and $\module m_q\in B_\out(C)$.  This contradicts the assumption that the component of $\module y$ in $C\setminus\{\module x\}$ is disjoint from $B_\out(C)$, thus proving part (\ref{it:i}).

  Now suppose $\module w\neq \module y$ is adjacent to $\module x$ along face $h$. Let $r$ be the cell adjacent to $\module x$ at $f^{\text{op}}$, and let $t$ be the cell adjacent to $r$ and $\module w$ not containing $\module x$. If $t$ is empty, then clearly $\module w\in B_\out(C)$. Otherwise, the module $\module m_t$ in cell $t$ is adjacent to $r$ (which is empty), so $\module m_t\in B_\out(C)$.  In either case, $\module w$ is in a component of $C\setminus\{\module x\}$ not disjoint from $B_\out(C)$, hence part (\ref{it:ii}) holds.

  Third, since $\module x$ is articulate in $C$, $\module x$ has degree at least $2$, so it is adjacent to at least one module $\module w\neq \module y$, proving part (\ref{it:iii}).

  Finally, let $e$ be a cell horizontally adjacent to a face $f^e$ of $\module x$. Assume $e$ is empty. Then let $e^+$ be the cell above $e$. Now suppose part (\ref{it:iv}) is false and $e^+$ is empty as well. 
  Then, $f^e$ must be adjacent to $f^\text{op}$, as all faces of $\module x$ except the one it shares with $\module y$ are adjacent to $f^{\text{op}}$. As $e$ and $e^+$ are both empty, $f^e$ and $f^{\text{op}}$ are slide-adjacent---and hence $f^e$ is also on the outer boundary of $C$. However, $f^e$ is also slide-adjacent, along its bottom edge, to some face of $\module y$ or a face of some neighbor $\module z$ of $\module y$. 
  In the former case, we have $\module y\in B_\out(C)$ which contradicts that $\module y$ is disjoint from $B_\out(C)$.
  In the latter case, we have $\module z\in B_\out(C)$ and, since $\module z$ is adjacent to $\module y$, the component of $C\setminus\{\module x\}$ containing $\module y$ is not disjoint from $B_\out(C)$, another contradiction. Thus, we have part (\ref{it:iv}).
\end{proof}

\begin{definition}\label{def:nearly-non-art}
  For a configuration $C$ of $n$ modules, a module $\module m$ on $B_\out(C)$ is said to be \term{nearly non-articulate} if $C\setminus\{\module m\}$ has exactly two connected components, one of which is disjoint from $B_\out(C)$.
\end{definition}

Our algorithm works by moving non-articulate modules on the boundary of $C$ and repeatedly stacking them on some module $\module s$. Therefore, we prove that we indeed can find some module $\module m \neq \module s$ on the boundary that either is non-articulate or can be made non-articulate. 

\begin{lemma}\label{lem:non-art-exists}
  For any configuration $C$ of $n\ge 2$ modules and a module $\module s\in B_\out(C)$, there is a module $\module m\in B_\out(C)$ with $\module m\ne\module s$ such that $\module m$ is either non-articulate or nearly non-articulate in $C$.
\end{lemma}

\begin{proof}
  As observed at the start of Section~\ref{sec:Structural_Properties}, 	$C$ contains at least two non-articulate modules; hence, $C$ has at least one non-articulate module $\module m_1\neq \module s$. If $\module m_1\in B_\out(C)$, then no further argument is required. Otherwise, suppose we have a set $M_{i-1} = \{\module m_1,\ldots,\module m_{i-1}\}\subset C\setminus B_\out(C)$ such that for each~$j$ with $1\le j < i$, the module $\module m_j$ is non-articulate in \mbox{$C\setminus\{\module m_1,\ldots,\module m_{j-1}\}$}. Then $C\setminus M_{i-1}$ is connected, so as before, $C\setminus M_{i-1}$ contains at least one non-articulate module $\module m_{i}\neq s$. Set $M_i = M_{i-1}\cup \{\module m_i\}$.

 For some minimal $t>1$, the module $\module m_t$ found in this way must be in $B_\out(C)$, as there are only finitely many modules in $C$. If $\module m_t$ is a non-articulate module of $C$, then we have 	the desired result. Otherwise, by the connectivity of $C\setminus M_t$, all of $B_\out(C)\setminus\{\module m_t\}$ lies in a single connected component of $C\setminus\{\module m_t\}$, so $\module m_t$ must have a neighboring cell not in $B_\out(C)$. Hence, $\module m_t$ must be adjacent to $\module m_i$ for some $1\le i < t$. By Lemma~\ref{lem:structure} with $\module x=\module m_t$, all modules not in the component of $\module m_i$ in $C\setminus\{\module m_t\}$ are in the component containing $B_\out(C)\setminus\{\module m_t\}$ (recall that $B_\out(C)$ is in a single component), thus removing $\module m_t$ leaves exactly two components one of which is disjoint from $B_\out(C)$. Hence, $\module m_t$ is nearly non-articulate, as required.
\end{proof}

\begin{definition}\label{def:closed}
    A set $F$ of cell faces is \emph{closed} if every edge of a face in $F$ is incident to an even number of faces in $F$.
\end{definition}

Using Definition~\ref{def:closed}, if $P$ is a set of cells, then the set of faces in $\partial P$ is closed. The following result shows the converse.  

\begin{lemma}\label{lem:boundary}
    Every bounded closed set of cell faces $F$ is the boundary of a bounded set of cells.
\end{lemma}
\begin{proof}
    Let $c$ be a cell and $\overrightarrow{r_c}$ be the vertical upwards ray from the center of $c$.
    We refer to the \emph{parity} of $c$ as the parity of the number of faces in $F$ that $\overrightarrow{r_c}$ intersects.
    Let $P_F$ be the set of all cells  $c$ with odd parity.
    Let $D$ be the symmetric difference between $\partial P_F$  and $F$ (viewing $\partial P_F$ as a set of faces).
    Notice that $D$ is closed since both $\partial P_F$  and $F$ are.
    We claim that $D$ has no faces parallel to the $xy$-plane.     
		For the sake of seeking a contradiction, assume face $f\in D$ is horizontal.
    If $f\in F$, then it is incident to two cells $c_1$ and $c_2$ with different parity. Then exactly one of the two is in $P_F$ and $f\in \partial P_F$, a contradiction.
    Similarly, $f$ cannot be in $\partial P_F$.
    
    We now claim that $D$ is empty. For contradiction, let $f$ be a highest face in $D$. Recall that $f$ must not be parallel to the $xy$-plane.
    Let $e$ be the top edge of $f$. By the choice of $f$, $e$ is not incident to a face in $D$ above it, and $D$ has no horizontal faces.
    Then, $D$ is not closed, a contradiction.
    We conclude that $\partial P_F=F$.

    It remains to show that $P_F$ is bounded. We show that $P_F$ is contained in the bounding box of $F$.
    For contradiction, let $c\in P_F$ be a cell below the bounding box.
    Let $p$ be the shortest path of adjacent cells connecting $c$ to a cell $c'$ which is not below the bounding box of $F$.
    Since $c'$ has even parity, $c'\notin P_F$, and $p$ must have crossed a face $f$ on the boundary of $P_F$.
    But $f\notin F$ since it lies below the bounding box.
    This is a contradiction because $\partial P_F=F$, as proven before.
\end{proof}

\begin{lemma}\label{lem:boundary-connected}
    Let $C$ be a connected configuration with a connected complement (where the connectedness of the complement is defined as in Section~\ref{sec:def}). Then, the slide-adjacency graph of $C$ is connected.
\end{lemma}

\begin{proof}
    Let $B$ be a component of the slide-adjacency graph of $C$.
    By the definition of slide-adjacency, $V(B)$ is closed, where $V(B)$ represents the vertices of $B$, i.e., a subset of boundary faces of $C$.
    Let $P_B$ be the set of cells described by Lemma~\ref{lem:boundary}, with $\partial P_B = V(B)$.
    We first claim that $P_B\subseteq C$. For contradiction, let $c$ be a cell in $P_B\setminus C$.
    Then $c\in \overline{C}$ and $\overline{C}\subset P_B$ because there are no boundary faces between adjacent cells in $\overline{C}$.
    Then $P_B$ is unbounded, a contradiction.

    We now observe that no cells in $P_B$ are adjacent to cells in $C$---as otherwise, the face between a pair of such cells would be in the boundary of $P_B\subseteq \partial C$, a contradiction.
    We then conclude that $P_B=P$, because $P$ is connected.
    Thus, $B$ is the entire slide-adjacency graph of $C$.
\end{proof}

In the following, we show that once a module reaches the outer boundary, it can reach any other position in the outer boundary without leaving it.

\begin{corollary}
    \label{cor:boundary-connected}
    Let $C$ be a connected configuration. The subgraph of the slide-adjacency graph of $C$ induced by the faces of the outer boundary of $C$ is connected.
\end{corollary}

During the execution of our algorithm, we need paths that a module can travel along while avoiding a particular face. The following result allows us to reroute a path in the slide-adjacency graph.

\begin{figure}[ht]
    \centering
    \includegraphics{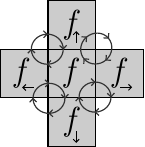}
    \caption{For a face $f$, with four neighbors $f_\uparrow$, $f_\rightarrow$, $f_\downarrow$, and $f_\leftarrow$, there are four cycles---one per vertex $v$---which visit $f$ and its two neighbors that are adjacent to $v$.}
    \label{fig:aroundF}
\end{figure}

\begin{lemma}\label{lem:avoid}
Given a face $f$ with two faces $f_1$ and $f_2$ which are slide-adjacent to $f$ there exists
a path from $f_1$ to $f_2$ in the slide-adjacency graph which does not use $f$. Moreover, such a path has length $O(1)$ for any fixed dimension $d$.
\end{lemma}
\begin{proof}
    First, note that all faces containing a vertex form a union of cycles, as each face can only be slide-adjacent to exactly one other face through an edge.

    Let $f_\uparrow$, $f_\rightarrow$, $f_\downarrow$, $f_\leftarrow$ be the four neighbors of $f$ in the slide adjacency graph (see Figure~\ref{fig:aroundF}). As $f$ has four vertices there are four cycles around $f$, each one contains $f$ and two of its neighbors (in rotation order). Hence w.l.o.g. there is a cycle that contains $f_\uparrow$, $f$, and $f_\rightarrow$. Therefore there is a path from $f_\uparrow$ to $f_\rightarrow$ excluding $f$. We can apply the same argument again to $f\rightarrow$, $f$, and $f_\downarrow$, and by composing the resulting paths, we can obtain a path from $f_\uparrow$ to $f_\downarrow$ avoiding $f$. Hence we can move from any $f_1$ to $f_2$ with one or two applications of this technique. Since the faces lie on the integer grid, each vertex is adjacent to $O(d)$ faces, so these cycles have length $O(d)$.
\end{proof}

The following is the primary technical tool that allows us to show that we can ``free'' a module on the boundary of any given configuration.

\begin{lemma}\label{lem:reconfigure}
  Given a configuration $C$ of $n\ge 2$ modules and a module $\module s\in B_\out(C)$, it is possible to reconfigure $C$ to a configuration $C'$, keeping $B_\out(C)$ fixed during the reconfiguration, so that $C'$ has a non-articulate module $\module x\neq \module s$ in $B_\out(C') = B_\out(C)$.
\end{lemma}

\begin{proof}
	We induct on $n$, the number of modules in $C$. When $n = 2$, both modules are non-articulate and must be in $B_\out(C)$, so this is trivially true even without reconfiguration. For the general case, we may find by Lemma~\ref{lem:non-art-exists} a module $\module x\in B_\out(C)\setminus\{\module s\}$ which is either non-articulate or nearly non-articulate in $C$.

 In the former case, $C=C'$, $\module x$ is the chosen module and we are done.

 In the latter case, $\module x$ is nearly non-articulate therefore $C\setminus \{\module x\}$ has exactly two components, the ``outer'' component $O$ and the ``inner'' component $I$; specifically, $I$ is the component disjoint from $B_\out(C)$, and \mbox{$O=C\setminus (I\cup\{\module x\})$}. By Lemma~\ref{lem:structure}(\ref{it:ii}), there is a unique module $\module y\in I$ adjacent to $\module x$. By Lemma~\ref{lem:structure}(\ref{it:iii}), there is a module $\module w\notin I$ adjacent to $\module x$, and by Lemma~\ref{lem:structure}(\ref{it:ii}) $\module w$ cannot be opposite from $\module y$. So, let $c$ be the cell adjacent to $\module y$ and $\module w$ not containing $\module x$. Cell $c$ must be empty since $\module w\notin I$. Let $f$ be the face of $\module y$ adjacent to cell $c$; it is clear that $f$ is on the outer boundary of $I$ (this is a direct consequence of Lemma~\ref{lem:structure}).

	Thus, since $I$ has fewer modules than $C$, the inductive hypothesis shows that we may reconfigure $I$ to $I'$ without moving $B_\out(I)$ and then find a non-articulate module $\module m\in B_\out(I')$ that is distinct from $\module y$. 
	It is then enough to show that we can move $\module m$ to a position where it is adjacent to both $O$ and $I$, making $\module x$ non-articulate.
  To do that, we find a path on the outer boundary of $I'\setminus\{\module m\}$ to $f$.
  If $\module m$ can indeed move along this path, then it can reach $f$ and, by its definition, we are done.
  Else, its movement is obstructed either by $\module x$ or by a module in $O$. We show that, at such position, $\module m$ becomes adjacent to both $O$ and $I'$. 
  In order to prevent a collision between $\module m$ and $\module x$, we make sure that the path that $\module m$ follows does not contain the face $f_{xy}$ shared by $\module x$ and $\module y$. 
  Note that $I'\setminus\{\module m\}$ is connected, and that by Corollary~\ref{cor:boundary-connected}, its outer boundary is connected in the slide-adjacency graph.
  Let $f_m$ be a face to which $\module m$ is attached, and note that $f_m$ is on the outer boundary of $I'\setminus\{\module m\}$.
  We can then find a path from $f_m$ to $f$.
  If such a path uses $f_{xy}$ we use Lemma~\ref{lem:avoid} to obtain the required path.
  We note that the fact that such path does not contain $f_{xy}$ is not enough to guarantee that the movement of $\module m$ does not intersect $\module x$.

  It remains to show that when $\module m$ cannot move along the computed path $\module m$ already connects $I'$ and $O$.
  Any two adjacent faces $g,g'$ on the path are connected at either a $90^\circ$ solid dihedral angle (as in Figure~\ref{fig:Cubemoves}(c)), a $180^\circ$ angle (as in Figure~\ref{fig:Cubemoves}(a)), or a $270^\circ$ angle.
  Assume that  $\module m$ is attached to $g$.
  If $g,g'$ form a $270^\circ$ dihedral angle, then $\module m$ is already attached to $g'$.
  If $g,g'$ form a $180^\circ$ dihedral angle, $\module m$ can perform a slide to attach to $g'$. If that position is occupied by a module in $O$ then $O$ and $I'$ are already adjacent, a contradiction.
  If that position is occupied by $\module x$ then either $g'=f_{xy}$ (contradicting the definition of the path) or $g'$ is a vertical face which would imply that $\module x$ has a neighbor in $I'$ different than $\module y$ (contradicting Lemma~\ref{lem:structure}(\ref{it:ii})).
  If $g,g'$ form a $90^\circ$ dihedral angle, $\module m$ can perform a rotation unless the edge shared by $g$ and $g'$ is pinched due to a module not in $I'$.
  If it is pinched due to a module in $O$, then $\module m$ is already adjacent to such a module.
  Similarly, if it is pinched due to $\module x$, then by Lemma~\ref{lem:structure}(\ref{it:iv}) $\module m$ is already adjacent to a module in $O$.
\end{proof}

\subsection{Reconfiguring into a Canonical Configuration}\label{sec:proof}
Using the results from above, we show any configuration $C$ can be reconfigured into a straight chain (a canonical configuration), proving the first part of Theorem~\ref{thm:main}. That is, any configuration $C$ can be reconfigured into another configuration $C'$. This result will be the basis of our algorithm that is presented in Section~\ref{sec:algorithm}. In the same section, we will also prove the second part of Theorem~\ref{thm:main} (i.e., a bound on the total number of moves required).

\begin{lemma}\label{lem:cannonical-reconfiguration}
    Any configuration $C$ can be reconfigured into a straight chain of modules (a canonical configuration).
\end{lemma}
\begin{proof}
Let $\module s\in B_\out(C)$ be a module with maximal $x_1$-coordinate, and let $f$ be the face of $\module s$ in the positive $x_1$-direction. Initially, denote $C_{0} = C$ and $Z_0=\varnothing$. We will iterate, maintaining the following invariants: After step $i-1$ ($1\le i\le n-1$), $\module s$ has not moved, and the configuration has the form $C_{i-1}\cup Z_{i-1}$, where $Z_{i-1}$ is a straight chain of $i-1$ modules emanating from face $f$ of $\module s$ in the positive $x_1$ direction, $C_{i-1}$ is connected, and $\module s\in B_\out(C_{i-1})$.

By Lemma~\ref{lem:reconfigure}, we may reconfigure $C_{i-1}$ to $C_{i-1}'$ while keeping $B_\out(C_{i-1})$ fixed in such a way that there is a module $\module x\in B_\out(C_{i-1}')$ different from $\module s$ that is non-articulate in $C_{i-1}'$. This implies that $\module x$ is non-articulate in $C_{i-1}'\cup Z_{i-1}$. By Corollary~\ref{cor:boundary-connected}, the subgraph of the slide adjacency graph induced by the outer boundary of $C_{i-1}' \cup z_{i-1}$ is connected so we may move $\module x$ along the boundary of $C_{i-1}'\cup Z_{i-1}\setminus\{\module x\}$ so that it extends the chain $Z_{i-1}$. Let $Z_{i}$ be this new chain of length $i$, and let $C_{i}$ be $C_{i-1}'\setminus \{\module x\}$. These clearly satisfy the required invariants.

After stage $n-1$, the reconfiguration is complete. 
\end{proof}

\section{\texorpdfstring{$\LocateAndFree$}{LocateAndFree}-based Algorithm}\label{sec:algorithm} %
\newcommand{\PostOrder}{\operatorname{\text{\textbf{PostOrder}}}}
The proof given in Section~\ref{sec:proof}
gives rise to a simple algorithm to reconfigure an $n$-module
configuration $C$ into a straight chain. Here we present this
algorithm (Algorithm~\ref{alg:V-to-chain}) and prove its correctness. We first require a recursive method that, given a configuration $C$
and a module $\module s\in B_\out(C)$ (along with a particular face of
$s$ on the outer boundary), modifies $C$ and returns a module $\module
x$ according to Lemma~\ref{lem:reconfigure}. We assume that each
module $\module m$ has previously been assigned a attribute
$\PostOrder(\module m)$ which sorts the modules of $C$ in the order of
finishing times of a depth-first search in $G_C$ beginning at $\module s$.

See Algorithm~\ref{alg:locandfree}, which converts
Lemma~\ref{lem:reconfigure} to a routine $\LocateAndFree$. Most of Algorithm~\ref{alg:locandfree} follows
Lemma~\ref{lem:reconfigure} directly. To prove
Algorithm~\ref{alg:locandfree} correct, we must address the comments
in lines~\ref{line:loc-nonart} and~\ref{line:loc-postorder}.

\begin{algorithm}
  \caption{Locate a module $\module x\in B_\out(C)$ satisfying
    Lemma~\ref{lem:reconfigure}, and reconfigure the interior of $C$
    to render $\module x$ non-articulate. Assumes $\PostOrder$ attributes
    in $C$ have been set.} %
  \label{alg:locandfree} %
  \begin{algorithmic}[1] %
    \STATE $\LocateAndFree(C,\module s) :=$ %
    \makeatletter\begin{ALC@g}\makeatother 
      \STATE Locate all faces in the outer boundary of $C$ by DFS from $\module s$. We obtain $B_\out(C)$. \label{line:loc-2}%
      \STATE Choose $\module x\in B_\out(C)$ with smallest $\PostOrder$. %
      \COMMENT {$\module x$ is (nearly)
        non-articulate}\label{line:loc-nonart}\label{line:loc-3} %
      \STATE Compute all modules in the component $O$ of
      \mbox{$C\setminus\{\module x\}$} containing $\module s$ by DFS.\label{line:loc-4} %
      \IF {$O$ contains all neighbors of $\module x$}\label{line:loc-5} %
      \RETURN{$\module x$}. %
      \ELSE %
      \STATE Let $\module y$ be the neighbor of $\module x$ in the other component~$I:=C\setminus (O\cup\{\module x\})$.\label{line:loc-8} %
      \STATE Let $\module m = \LocateAndFree(I,\module y)$.\label{line:loc-9} %
      \COMMENT{Use existing $\PostOrder$
        labels.}\label{line:loc-postorder} %
      \STATE Move $\module m$ to connect $O$ and $I$ as in
      Lemma~\ref{lem:reconfigure}, locating its path by DFS across the outer boundary of \mbox{$C\setminus\{\module m\}$}.\label{line:loc-10} %
      \RETURN{$\module x$}. %
      \ENDIF %
      \makeatletter\end{ALC@g}\makeatother %
    \STATE \textbf{end LocateAndFree} %
  \end{algorithmic}
\end{algorithm}

First, we must show that the module $\module x\in B_\out(C)$ with
minimal $\PostOrder$ is non-articulate or nearly non-articulate.  If
$\module x$ is articulate in $C$, then a path from $\module s$ to any
module $\module t\not\in O$ must pass through $\module x$, meaning
 {
    \begin{equation*}
        \PostOrder(t)\le\PostOrder(x).
    \end{equation*}
}

This means $\module t$ cannot be in $B_\out(C)$, by the minimality of
$\PostOrder(\module x)$. Thus, any connected component of
\mbox{$C\setminus\{\module x\}$} not containing $\module s$ is disjoint from
$B_\out(C)$, so Lemma~\ref{lem:structure} applies, proving that $\module
x$ is indeed nearly non-articulate.

We must also prove that the attribute $\PostOrder$ sorts the modules of
$I$ in a post-order from $\module y$. By choice of $\module x$, the original
depth-first tree restricted to $I$ must itself be a valid depth-first
tree of $I$ rooted at $\module y$, and thus the $\PostOrder$ field is
correctly sorted, as needed.

Now we present Algorithm~\ref{alg:V-to-chain}, which rearranges $C$
into a straight chain, using Algorithm~\ref{alg:locandfree} as a subroutine.  The proof of correctness of Algorithm~\ref{alg:V-to-chain} follows directly from
the results in Section~\ref{sec:proof}.

\begin{algorithm}[!ht]
  \caption{Reconfigure $C$ into a straight chain $\{\module s\}\cup Z_{n-1}$.}
  \label{alg:V-to-chain}
  \begin{algorithmic}[1]
    \STATE Fix a module $\module s\in C$ with maximal $x_1$-coordinate. %
    \STATE Set $C_0 = C$ and $Z_0 = \varnothing$. %
    \FOR{$1\le i < n$}\label{algstep:for} %
    \STATE Set the $\PostOrder$ fields with a depth-first search
    rooted at $\module s$. %
    \STATE Define $\module x := \LocateAndFree(C_{i-1},\module s)$.  %
    \STATE By depth-first search across the outer boundary faces of
    $C_{i-1}\cup Z_{i-1}\setminus\{\module x\}$, move $\module x$ to extend
    $Z_{i-1}$. Define $C_i := C_{i-1}\setminus\{\module x\}$ and $Z_i = Z_{i-1}\cup\{\module x\}$. %
    \ENDFOR %
  \end{algorithmic}
\end{algorithm}

\subsection{Algorithm Analysis}\label{sec_error}

We first briefly comment on the analysis from \cite{abel2008universal}: Each of the $n-1$ iterations of Algorithm~\ref{alg:V-to-chain} may in principle make $O(n)$ recursive calls to $\LocateAndFree$. In each recursive call, $\module m$ moves $O(n)$ times; thus, the overall number of moves is $O(n^3)$. Relative to \cite{abel2008universal}, we are more careful in how we define the path in the proof of Lemma~\ref{lem:reconfigure}. In~\cite{abel2008universal}, Abel and Kominers use a similar induction on $C$, taking the interior component $I$ of $C$, and reconfiguring it into $I'$; they then define their path on the outer boundary of $I'$. The Abel and Kominers~\cite{abel2008universal} analysis gave no proof of connectivity of the outer boundary of $I'$. Additionally $\module x \notin I'$, so the face $f_{yx}$ that $\module y$ shared with $\module x$ is on the boundary of $I'$ even though it is actually inaccessible as $\module x$ is attached to $f_{yx}$, so this path would need to avoid collision with $\module x$. Finally, since the path is defined on the outer boundary of $I$ and not $I'\setminus \{\module m\}$, it would be possible that the path found uses a face adjacent to the initial position of $\module m$, which no longer exists as $\module m$ is the mobile module, hence the path needs to avoid $\module m$'s initial position as well. The latter two issues were also unaddressed in~\cite{abel2008universal}.

We now present an improved analysis of the algorithm, addressing the connectivity issue along the way. We show that although $\LocateAndFree$ may recursively call itself $O(n)$ times, it only uses $O(n)$ moves over the entire execution of the algorithm. Therefore we can use $\LocateAndFree$ to make a single module on the boundary non-articulate with $O(n)$ moves. For this, we need to assume that the path computed in Lemma~\ref{lem:reconfigure} is the \textit{shortest} path from $f_m$ to $f$ while avoiding $f_{xy}$---this can be computed with breadth-first search after deleting $f_{xy}$ from the slide adjacency graph of $I'\setminus\{\module m\}$.

\begin{figure}[htb]
    \centering
    \includegraphics[scale=.3]{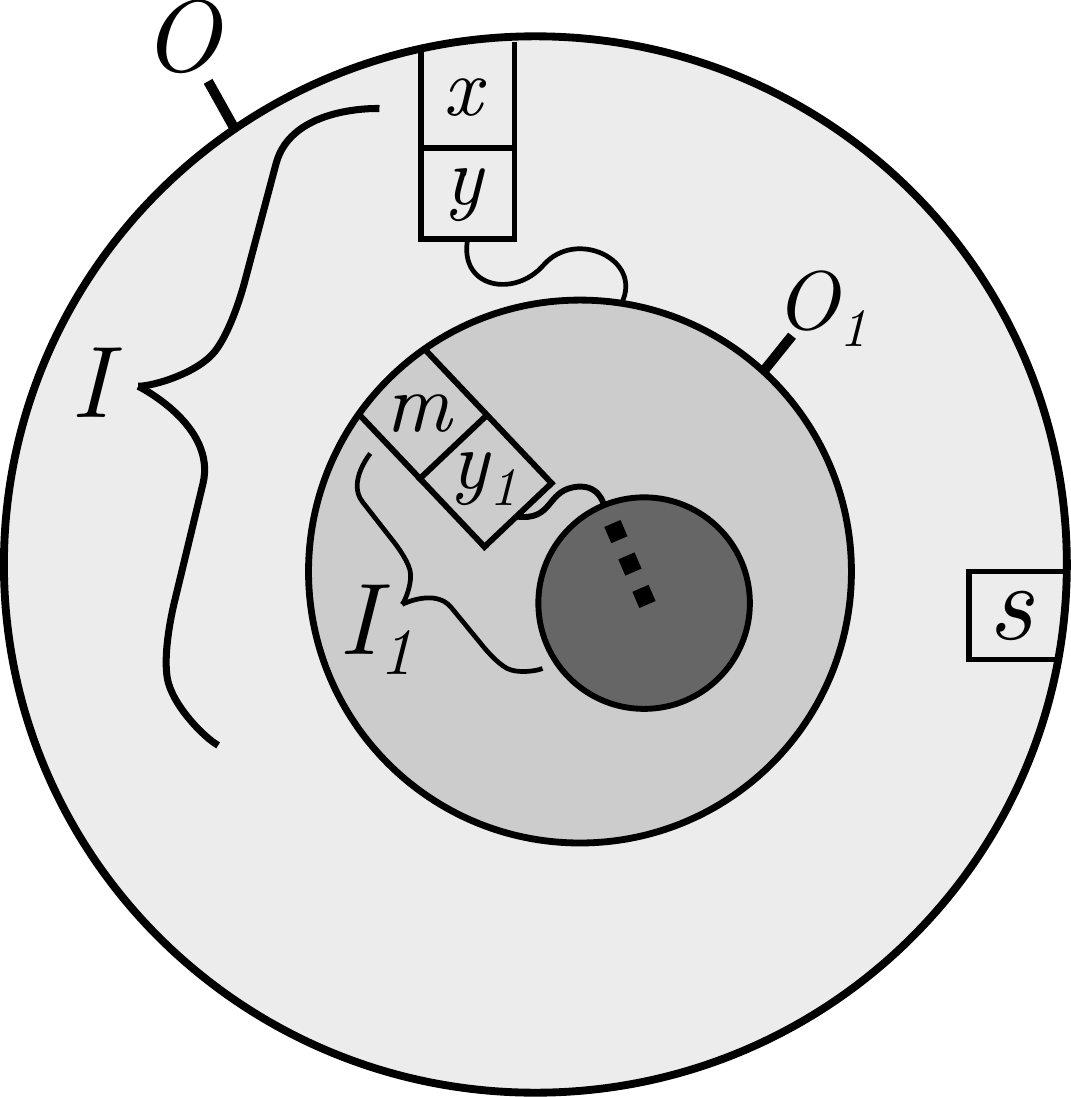}
    \caption{$O_i$ and $I_i$ are connected by one module; $O_{i+1}$ and $I_{i+1}$ are a subset of $I_{i}$.}
    \label{fig:IandONested}
\end{figure}

\begin{lemma}\label{lem:improv}
    $\LocateAndFree($C$,\module s)$, with all its recursive calls, executes $O(|I|)$ moves in any fixed dimension, where $|I|$ is the number of modules in the configuration $I$ as defined in Algorithm~\ref{alg:locandfree}.
\end{lemma}
\begin{proof}
    We use induction on the number of recursive calls. 
    If $\LocateAndFree($C$,\module s)$ does not make any recursive call, then no move is performed.
    We consider the recursive call $\LocateAndFree(I,\module y)$ in line 9 of Algorithm~\ref{alg:locandfree}.
    Let $O_1, I_1, \module x_1 = \module m$ and $\module m_1$ play the role of $O, I, \module x$ and $\module m$ for the recursive call, respectively. Thus $O_1\cup I_1\cup \{\module m\}=I$.
    By inductive hypothesis, $\LocateAndFree(I,\module y)$ performs $O(|I_1|)$ moves, transforming $I$ into a configuration $I'$ with the same number of modules.
    It indeed only changes $I_1$ into $I_1'$, maintaining $O_1$ by construction.
    It is then enough to show that $\module m$ moves at most $O(|O_1|)$ times. The general structure of $O, I, O_1, I_1, \dots$ are depicted in Figure~\ref{fig:IandONested}.

    Recall that $\module m$ moves on the outer boundary of $I'\setminus\{\module m\}$. Thus it moves on the outer boundary of $O_1\cup I_1'$.
    By choice of $\module x_1=\module m$, $f$ lies on the boundary of $O_1$.
    By Lemma~\ref{lem:structure}(\ref{it:iii}), if $|I_1|\neq 0$, $m$ is adjacent to a module in $B_\out(O_1)$ of which one face is $f_m$.
    Without loss of generality, assume the top face of $\module m$ is on the outer boundary of $I'$.
    Note that $B_\out(I')  = B_\out(O_1\cup\{\module m\}\cup I_1')= B_\out(O_1\cup\{\module m\})$, i.e., $I_1'$ is completely ``inside'', only getting exposed when we delete $\module m$. 
    Thus,
    the four faces that are slide-adjacent to the top face of $\module m$ are the only faces 
    of $\partial O_1$ that are slide-adjacent to faces of $\partial I_1'$ in the outer boundary of $O_1\cup I_1'$.
    Note that $f_m$ is slide adjacent through its top edge to at least one of these four faces.
    The path taken by $\module m$ might use faces of $I_1'$ but it eventually enters $O_1$ for the last time to reach $f$, thus passing through one of these four faces.
    By Lemma~\ref{lem:avoid}, the distance between these four faces is $O(d)$.
    Thus, the path uses $O(d)$ faces of $I_1'$, and the path length is $O(|O_1|)$ given constant dimension.
\end{proof}

\begin{proof}[Proof of Theorem~\ref{thm:main}]
    Lemma~\ref{lem:cannonical-reconfiguration} proves that universal reconfiguration between any two connected configurations of modules in three dimensions is possible. Thus, to complete the proof of Theorem~\ref{thm:main} it suffices to bound the number of moves. 
		
		Essentially, the algorithm requires a linear number of calls to $\LocateAndFree$. By Lemma~\ref{lem:improv}, each call requires $O(n)$ moves. Additionally, we move each module so that it forms a line with other modules we previously obtained with $\LocateAndFree$. We can then bound the number of moves used to put each module in its canonical position by $O(n)$. 
This gives a total of $O(n^2)$ moves as claimed. 
\end{proof}

\section{Bounded-Space Algorithm}
\label{sec:inPlace}

In this section, we use $\LocateAndFree$ to design a different reconfiguration algorithm with two significant properties. First, the algorithm is \emph{in-place}---during the whole reconfiguration process, modules will be contained in the bounding box of the union of the start and end configuration (plus possibly a small $O(1)$ margin). Second, the algorithm  is \emph{input-sensitive}---the number of moves needed is bounded by the number of modules $n$ and the volume of the start and end configurations.

Let $B$ be the bounding box of the union of both the start and end configurations with dimensions $x_M \times y_M \times z_M$. By translation and coordinate reflection, we can assume that one of the corners of $B$ is the origin and that its opposite corner is $(x_M, y_M, z_M)$. Moreover, by renaming the axis, we can also assume that $2 \leq x_M \leq y_M \leq z_M$. With these definitions, we can state the Bounded-Space algorithm:

\begin{theorem}\label{thm:in-place}
  Given any two connected unlabeled configurations $C$ and $C'$ each having $n\ge 2$ modules, there exists an in-place reconfiguration of $C$ into $C'$ using $O(n\cdot\min\{n,x_My_M + z_M\})$ sliding moves.
\end{theorem}

It is easy to find examples where $x_M$, $y_M$, and $z_M$ range between constant and linear. If $C$ and $C'$ span the same amount in all dimensions and are dense (i.e., the number of occupied positions is similar to the number of empty positions), then $x_M \approx y_M \approx z_M$ and the algorithm will reduce the number of moves by a factor of roughly $\approx n^{1/3}$. The improvement can grow up to a factor $\sqrt{n}$ when $x_M \approx y_M \approx \sqrt[4]{n}$ and $z_M \approx \sqrt{n}$. In the general case, we cannot guarantee that the number of moves will go down, yet we can ensure that algorithm is now in-place.

We note that the description below does not technically follow the \emph{in-place} model introduced by Moreno and Sacrist\'an~\cite{moreno2020reconfiguring} (we consider the bounding box of the union of both configurations whereas their algorithm restricts itself to the union of both bounding boxes). In a later section we will explain how to modify our algorithm to work in their more typical, in-place model.

\subsection{Algorithm Overview}
Similar to the algorithm described in Section~\ref{sec:algorithm}, the Bounded-Space algorithm reconfigures the initial configuration to some canonical form, and then follows the same steps in reverse to reach the target configuration. For simplicity of its description, we assume that $x_M$ is odd and $y_M$ is even (or the reverse). We further assume that $n > x_My_M+2z_M $. We show how to remove these assumptions in Section~\ref{sec_details}.

We now define our canonical form. Intuitively, we want to compress all blocks forming a parallelogram of low $z$ coordinate. However, since $n$ may not be a multiple of $x_My_M$ ($n$ may even be a prime number) a parallelogram may not be possible.

\begin{figure}[htb]
    \centering
    \includegraphics{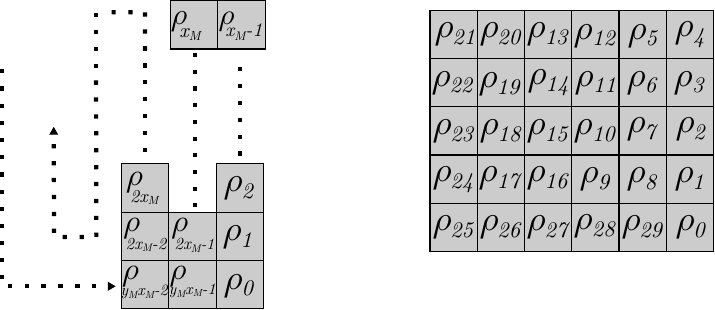}
    \caption{A Hamiltonian cycle for a layer---left is a generic pattern; right is a concrete example on a $5\times 6$ grid.}
    \label{fig:HamPath}
\end{figure}

By our assumption of parity, we know that there is a Hamiltonian cycle on the grid graph of size $x_M\times y_M$. Fix any such cycle and virtually walk the Hamiltonian cycle starting from the origin. Number all positions in the grid in increasing order as traversed through the cycle. Let $\rho_0, \ldots \rho_{x_My_M-1}$ be the ordering created, see Figure~\ref{fig:HamPath}. We identify $\rho_i$ with the $2$-dimensional coordinates of the corresponding grid vertex. We do a slight notation abuse and use $(\rho_i,z')$ to denote the $3$-dimensional position corresponding to lifting point $\rho_i$ to the plane $z=z'$.

\begin{figure}[ht]
    \centering
    \includegraphics{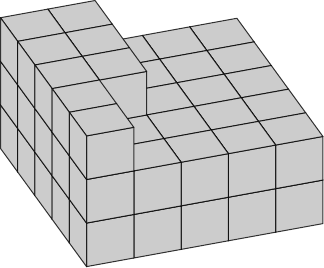}
    \caption{The compact configuration for 58 modules and bounding box $5 \times 5 \times z$ for all $z \geq 3$.}
    \label{fig:Compact}
\end{figure}

Suppose there is some module at position  $(\rho_{i},z')$ for some $0\leq i\leq x_My_M-1$ and $0\leq z' \leq z_M$. We say that a module is \emph{necessary} in either of the following conditions:
\begin{itemize}
    \item $i = 0$ and $z' = 0$ (i.e., module $(\rho_0,0)$ is always necessary).
    \item $ i = 0$, $z' > 0$ and position $(\rho_{x_My_M-1},z'-1)$ is occupied.
    \item $i > 0$ and $(\rho_{i -1 },z')$ is occupied. 
\end{itemize}

If all modules in a configuration are necessary we say that the configuration is \emph{compact}. In other words, before position $(\rho_i, z)$ can be occupied we must occupy previous positions in the same plane ($(\rho_0,z)\dots (\rho_{i-1},z)$) as well as all other positions of lower $z$ coordinate.

The Bounded-Space algorithm has three distinct phases executed one by one in order:
\begin{description}
    \item[Scaffolding Phase] Use $\LocateAndFree$ to construct a structure (``scaffolding'') that will help move other modules.
    \item[Meld Phase] With the scaffolding, shift all modules down until they hit another module. The result will be a configuration that is almost compact (most modules will be necessary).
    \item[Reconfiguration to Compact form] Move clumps of modules downwards one at a time until the configuration ends in compact form.
    
\end{description}

\subsection{Scaffolding Phase}

\begin{figure}[ht]
    \centering
    \includegraphics[scale=.25]{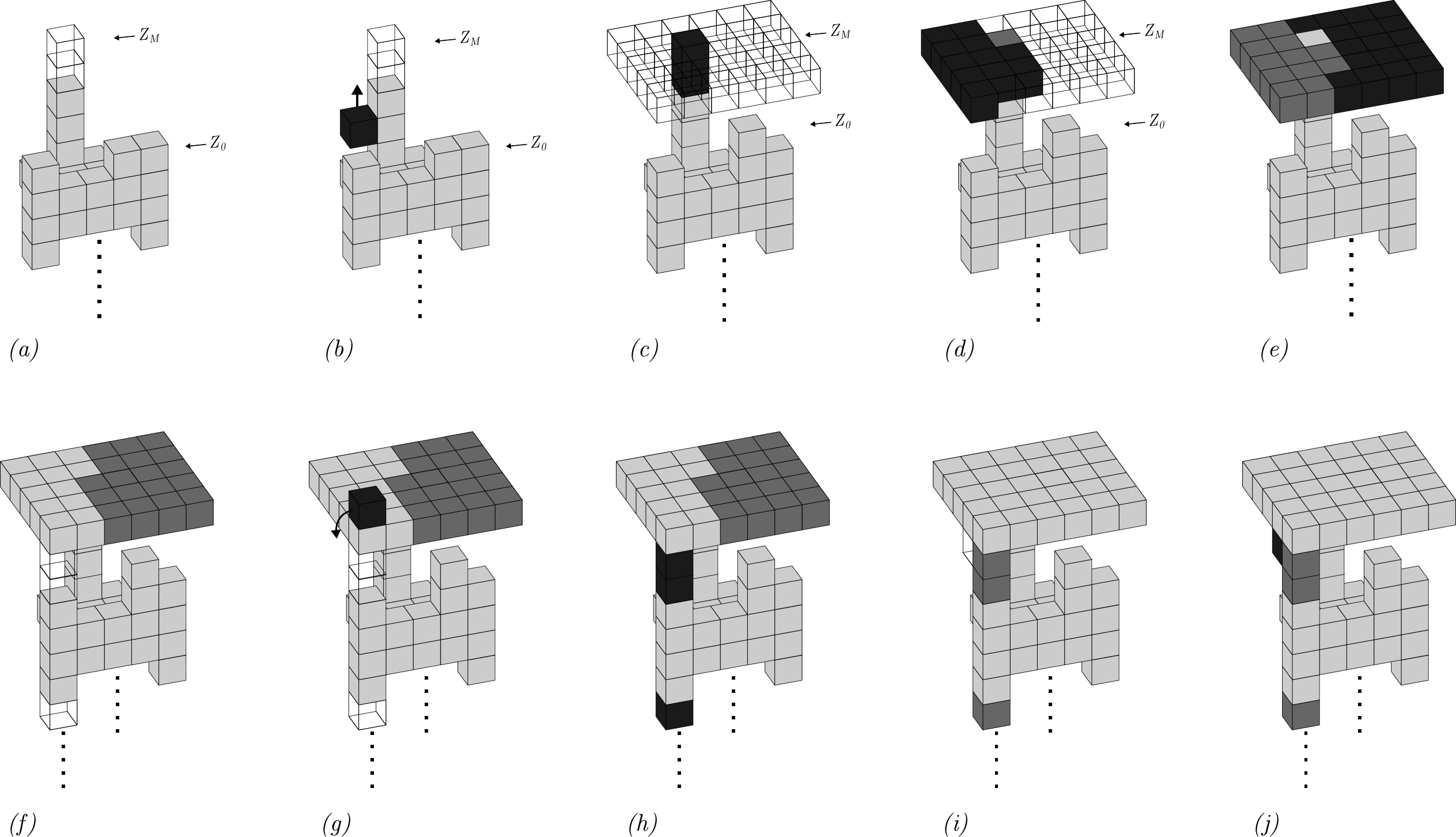}
    \caption{The Scaffolding Phase. Images (a-c) show filling up to $(\rho_{i_0},z_M)$; (d) shows the filling of positions $(\rho_j,z_M)$ where $j < i_0$; (e) shows filling for $j > i_0$; (f-h) show filling all $(\rho_0, Z)$; and (i-j) show filling the line $(\rho_0,z')$ for $z'\leq z_M$.}
    \label{fig:scaffolding}
\end{figure}

The goal of the scaffolding phase is to form a structure that will allow other modules to move without the use of $\LocateAndFree$. The first step of this phase is to fill in the intersection between plane $z=z_M$ and $B$ (in other words, fill in the rectangle $(x,y,z_M)$ for all $x\leq x_M$ and $y\leq y_M$). This can be done in a fairly straightforward way:

\begin{itemize}
\item Let $z_0$ be the greatest $z$-coordinate of a position occupied in the starting configuration. Let $i_0$ be the smallest index such that $(\rho_{i_0},z_0)$ is occupied. If $z_0=z_M$ we move onto the next step. Otherwise we call $\LocateAndFree$ $z_M-z_0$ times, placing freed modules in positions $$(\rho_{i_0},z_0+1), (\rho_{i_0},z_0+2), \ldots$$ until $(\rho_{i_0},z_M)$ is occupied. 

\item At this point we know that $(\rho_{i_0},z_M)$ is occupied by a module and no position $(\rho_j,z_M)$ is occupied (for $j<i$). Continue calling $\LocateAndFree$ this time placing modules in positions $(\rho_{i_0-1},z_M), (\rho_{i_0-2},z_M), \ldots $ until position $(\rho_0,z_M)$ is occupied. Note that this movement is always possible (move the module to position $(\rho_{j+1},z_M+1)$ and then do a rotation into $(\rho_{j},z_M)$). In this stage, exactly $i_0-1$ many calls to $\LocateAndFree$ will be needed.

\item Next we use $\LocateAndFree$ in a symmetric fashion to fill in positions $(\rho_{i_0+1},z_M)$, $(\rho_{i_0+2},z_M), \ldots (\rho_{x_My_M -1},z_M)$. However, we can no longer guarantee all positions are empty (this would only happen if $z_0 = z_M$). If a position is already occupied, simply skip it and place the module in the next empty position in the sequence.

\item The next goal is to fill in positions $(\rho_{0},z_0-1), (\rho_{0},z_0-2), \ldots (\rho_{0},0)$. We call $\LocateAndFree$ to free a module and move it into the first unoccupied position in this sequence. This position can be reached by the module as follows: move to position $(0,0,z_M +1)$, do a rotation into position $(-1,0,z_M)$, and perform slide operations downward until no longer possible as $(0,0,z')$ is occupied but $(0,0,z' -1)$ is not for some $0< z' <z_0$. Complete the operation with a rotation into $(0,0,z')=(\rho_0,z')$. Repeat until all positions including $(\rho_{0}, 0)$ are occupied.

\item Finally, if position $(\rho_{1},z_M-1)$ is unoccupied, we use $\LocateAndFree$ one last time and place that module in that position. This additional module will be used to help other modules move. We stick to the literature convention~\cite{a.akitaya_et_al:LIPIcs.SoCG.2021.10} and call this helping module a \emph{musketeer}.
\end{itemize}

\begin{lemma}
The Scaffolding phase will require at most $x_My_M +2z_M-1$ calls to $\LocateAndFree$. Moreover, once a module reaches the outer boundary, it will perform at most $O(n)$ additional moves. 
\end{lemma}
\begin{proof}
The bound in the number of calls follows from the fact that we fill in a plane, two lines (restricted within the bounding box), and an additional position. The bound on the additional number of moves per module follows from the fact that the longest a module may move is move up one of the lines, walk over the plane and then walk down the other line.
\end{proof}

\subsection{Meld Phase}
Once the scaffolding phase has finished, the top layer is fully occupied by modules (we say that those modules form the \emph{slab}). In the meld phase, we alternate between a phase in which we move the modules in the slab ``down'' (decreasing their $z$ coordinate) and a cleanup phase (in which we ensure structural properties are preserved). These alternate until the slab reaches the $z=0$ plane. The invariant of this phase is that modules below the slab remain unchanged, whereas all modules above are clumped together and will be easy to move in the third phase (exact definition below). 

\paragraph*{Lowering the Slab}

In our algorithm we need to track the position of the slab. When we say that the slab is at $z$-coordinate $z_i$ we mean that position $(x,y,z_i)$ is occupied for all $x\leq x_M$ and $y\leq y_M$. Since we do not place constraints on the initial configuration it is possible that this condition is satisfied at the same time at different coordinates. We only care about the slab we filled during the scaffolding phase (the slab at $z_m$) and track how it is lowered down one unit at a time. Before we lower the slab we require the following invariants:
\begin{enumerate}
	\item The slab is at $z_i$ for some $z_i \geq 0$
	\item Positions $(\rho_0,z')$ are occupied for all $z' \leq z_M$.
	\item If $z_i > 0$, position $(\rho_1,z_i-1)$ is occupied.
	\item If position $(\rho_j,z')$ is occupied for some $z' > z_i$ and $j>0$, then the module is necessary.
\end{enumerate}

Whenever these invariants are satisfied we say that the slab is \emph{ready to drop}. Note that, by construction, the Scaffolding phase ensures that the slab at $z_i=z_M$ is ready to drop: the first three conditions are satisfied by design and the fourth condition is true because $z_M$ is the maximal $z$ coordinate in $B$. In order to lower the slab we lower positions $(\rho_0,z_i-1), (\rho_1,z_i-1), \ldots$ one by one in order. Once all positions have been lowered we can say that the slab has been lowered. Say we focus on lowering module $(\rho_j,z_i)$ for some $j \geq 2$ (cases $j=1,2$ will be handled afterwards). Position $(\rho_j,z-i)$ is \emph{ready to drop} if the following properties are satisfied: 
\begin{itemize}
    \item Positions $(\rho_j,z_i-1)$ are occupied for all $j<i$. This is because we lower modules in that specific order.
    \item Positions $(\rho_j,z_i)$ are occupied for all $j \geq i$. Again, this is guaranteed by the order in which we lower modules (these positions have not been considered yet).
    \item Position $(\rho_{i-1},z_i)$ is occupied by the musketeer module.
\end{itemize}

Initially, The scaffolding phase guaranteed that positions $(\rho_0,z_i)$, $(\rho_1,z_i)$, $(\rho_0,z_i-1)$ and $(\rho_1,z_i-1)$ are occupied: the first two are occupied because of the slab, the third was occupied when we filled the vertical segment $(\rho_0,i)$ and the last position was occupied in the very last step of the scaffolding phase. We assign the musketeer role to the module at $(\rho_1,z_i)$, which implies that $(\rho_2,z_i)$, is \textit{ready to drop}.

\begin{figure}[htb]
    \centering
    \includegraphics[scale=.7]{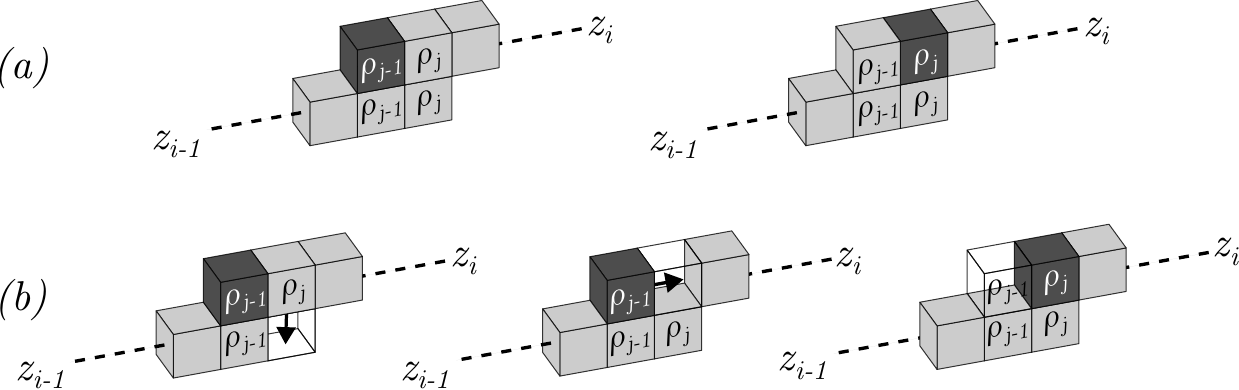}
    \caption{If $(\rho_j,z_i -1)$ is occupied, as in (a), then we do not perform any moves and simply change the module doing the role of musketeer (denoted in dark gray); otherwise, as in (b), we perform two slide operations and the same module remains as the musketeer.}
    \label{fig:musketeer}
\end{figure}

We consider two cases when module $(\rho_j,z_i)$ is ready to drop for some $j>1$ (see Figure~\ref{fig:musketeer}):
\begin{description}
    \item[Position $(\rho_j,z_i-1)$ is occupied] We don't have to do anything as the position we want to occupy is already occupied. The module at $(\rho_j,z_i)$ becomes the musketeer (the previous module that was a musketeer stays in the same place and becomes a regular module). Note that with both $(\rho_j,z_i)$ and $(\rho_j,z_i-1)$ this guarantees that $(\rho_{j+1},z_i)$ is ready to drop.
    
    \item[Position $(\rho_j,z_i-1)$ is empty] In this case we have a $2\times 2 \times 1$ configuration in which 3 positions are occupied ($(\rho_{j-1},z_i)$, $(\rho_{j-1},z_i-1)$ and $(\rho_j,z_i)$) and the last one ($(\rho_{j},z_i-1)$)is empty. Thus, with a slide operation, we can move the module at $(\rho_j,z_i)$ one unit down. Next, slide the musketeer from $(\rho_{j-1},z_i)$ into $(\rho_{j},z_i)$. This also guarantees that $(\rho_{j+1},z_i)$ is ready to drop.
\end{description}

\begin{lemma}
After at most $2x_My_M -4$ slide operations, we can lower the slab by one coordinate. Moreover, none of the operations performed will break the connectivity of the configuration.
\end{lemma}
\begin{proof}
The bound on the number of slides follows from the fact that we lower $x_My_M -2$ many modules (positions $(\rho_0,z_i)$ and $(\rho_1,z_i)$ are skipped) and each lowering requires at most 2 slide operations.

Connectivity follows from the fact that we lower the modules in order of a Hamiltonian cycle. Say that $(\rho_j,z_i)$ is ready to drop. If $(\rho_j,z_i-1)$ is occupied we do not do any moves, so clearly connectivity is not an issue. If $(\rho_j,z_i-1)$ is empty, the modules of the slab are split into two groups: the ones that have already been dropped (that is, modules $(\rho_0,z_i-1), (\rho_1,z_i-1), \ldots (\rho_j-1,z_i-1)$) and the ones that have not yet been dropped (modules $(\rho_{j},z_i), (\rho_{j+2},z_i), \ldots (\rho_{x_My_M-1},z_i)$) ), see Figure~\ref{fig:SlabLowering}. In addition to those modules, we know that positions $(\rho_0,z_i)$ and $(\rho_{j-1},z_i)$ are occupied as well (the former was guaranteed during the scaffolding phase and the latter is occupied by the musketeer). All the modules mentioned above form a cycle, thus the first slide operation is possible. The second slide operation ensures that the cycle is formed again, and that $(\rho_{j},z_i)$ is ready to drop. 
\begin{figure}[ht]
    \centering
    \includegraphics[scale=.5]{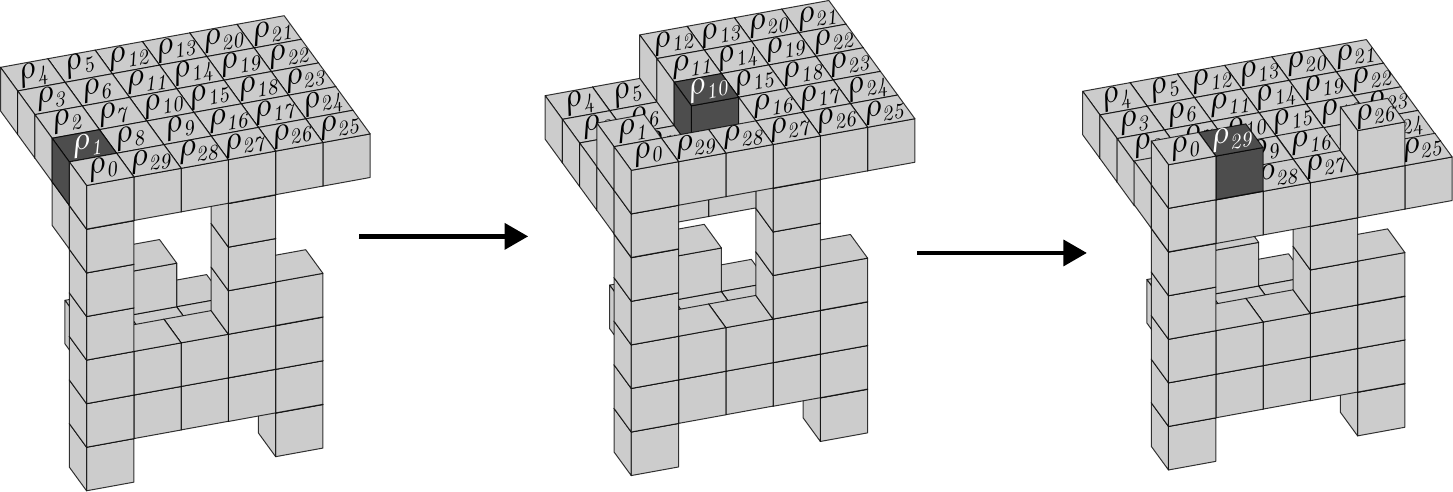}
    \caption{The process of lowering the slab. (left image): Initial position, $(\rho_2,z_i)$ is ready to drop; module at $(\rho_1,z_i)$ is assigned the musketeer role. (middle): $(\rho_{10},z_i)$ has been lowered; that position is now occupied by the musketeer and $(\rho_{11},z_i)$ is ready to drop. (right): Once the slab has been lowered, some modules will remain immediately above the new slab coordinate; none of these modules is a cut vertex.}
    \label{fig:SlabLowering}
\end{figure}
\end{proof}

After we have lowered every single position in the slab we can say that the slab has been lowered and decrease $z_i$ by one. Note that some modules may be left in the previous position of the slab. However, it is important to notice that none of those modules can be articulate. This is trivially true when $z_i=z_M$ (i.e., we lowered the slab for the first time). In general, this is true because of invariants 3 and 4 (most modules above the slab are necessary and thus connected to the rest of the configuration via the line of modules at $(\rho_0,z')$). 

\paragraph*{Cleanup Phase}
Once the slab has been lowered and the value of $z_i$ has been updated, some positions in $z_i+1$ may still be occupied:

\begin{itemize}
\item $(\rho_0,z_i+1)$ will be since this module is never moved nor does it become the musketeer
\item Position $(\rho_{x_My_M-1},z_i+1)$ is occupied by the musketeer
\item Additional positions in the $z_i+1$ may be occupied based on the number of occupied positions in the $z_i$ plane before the slab was lowered (although the modules may have moved while having the musketeer role).
\end{itemize}

Before we can start the next lowering of the slab, we need to ensure that position $(\rho_1,z_i-1)$ is occupied (it is necessary to ensure that $(\rho_2,z_i)$ is ready to drop). So, in the cleanup phase for $z_i$ we move the musketeer at $(\rho_{x_My_M-1},z_i+1)$ to $(\rho_1,z_i)$. This step is skipped if position $(\rho_1,z_i)$ was already occupied (or $z_i=0$).

Say that now $k>1$ many positions in the $z=z_i+1$ plane are occupied. We compress these modules to occupy positions $(\rho_0,z_i+1), \ldots (\rho_{k-1},z_i+1)$. This is done by sliding modules over empty positions in the Hamiltonian cycle to positions of smaller index (i.e., each time there exists an index $j$ such that $(\rho_{j+1},z_i+1)$ is occupied and $(\rho_{j},z_i+1)$ is not, we perform a slide operation). The order in which we execute the slide operations is irrelevant. 

\begin{lemma}
After the cleanup phase, the slab at $z_i$ is ready to drop. Moreover, the total number of module moves performed over all cleanup phases is bounded by $O(n(x_M+y_M))$. 
\end{lemma}
\begin{proof}
The first of the four invariants we require is guaranteed in the Meld phase when we lower the slab. The second one remains true during the whole process because positions $(\rho_0,z')$ for all $z'\geq 0$ start occupied and those modules never become musketeers (so they never move in either phase). The third and fourth invariant are directly ensured in the cleanup phase.

It remains to show the bound on the number of moves: moving the musketeer from $(\rho_{x_My_M-1},z_i+1)$ to $(\rho_1,z_i)$ requires a constant number of moves (2 slides and 3 rotations to be precise). Once a module stops being a musketeer it can never regain the role and will slide over the Hamiltonian graph exactly once. During this phase it can only move $x_My_M-1$ many positions. Overall, we have that a single module will never perform more than $O(1+x_My_M)$ many moves, proving the Lemma.
\end{proof}

\subsection{Reconfiguration to Compact form}\label{sec_almostcompact}

\begin{figure}[ht]
    \centering
    \includegraphics[scale=.75]{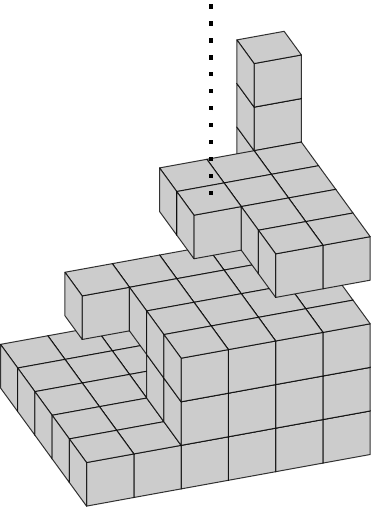}
    \caption{After the meld phase, we have an almost compact configuration.}
    \label{fig:clumps}
\end{figure}

Once the Meld phase has ended most modules will be in compact form: the only exception can be modules in the line $(\rho_0,z')$ (see Figure~\ref{fig:clumps}). This last phase will move modules around to ensure all modules are necessary. 

We do so by looking at the symmetric difference: any module that is in the current configuration and not in the compact configuration is an \emph{excess} module. When the reverse condition is held we say that the position is \emph{lacking}. In this final phase we need a one-to-one mapping between excess modules and lacking positions in a way that the number of moves is bounded and connectivity is never an issue.

We say that position $(\rho_i,z_j)$ comes after $(\rho_{i'},z_{j'})$ 
if an only if either $(a)$ $z_j>z_{j'}$ or $(b)$ $z_j=z_{j'}$ and $i>i'$. This is a total ordering on the positions of the box. Our mapping matches the largest excess module and moves it to the with the smallest lacking position. Once this is done, the symmetric difference has decreased by two and we recurse until the two configurations are equal. 

\begin{lemma}
The largest excess module can always move to its mapped lacking position in $O(x_My_M +z_M)$ moves. 
\end{lemma}
\begin{proof}
Say we are considering moving the the largest excess module $m$ occupying position $(\rho_i,z_j)$. First notice that any module at position $(\rho_{i'},z_{j'})$ is connected in $G_C$ to position $(\rho_0,0)$ over positions $(\rho_{i'-1},z_{j'}), (\rho_{i'-2},z_{j'}), \ldots, (\rho_{0},z_{j'})$ and then following the line $(\rho_{0},z_{j'}-1)$, $(\rho_{0},z_{j'}-2), \ldots (\rho_0,0)$. In particular, the largest excess module $m$ never appears in any of the paths to the origin. This implies that moving $m$ will not jeopardize the connectivity of the configuration. 

By the maximality of the excess module we either have that either $(\rho_{i+1},z_j)$ is empty or $i=x_My_M-1$. In either case we can move it within the plane $z_j$ until $m$ occupies or is adjacent to $(\rho_{0},z_j)$. Once at that position we can lower its $z$ coordinate by sliding over the modules of the line $(\rho_{0},z')$ until we reach the desired coordinate. We reach the final desired position by sliding on the existing modules in that plane. The desired position will be reachable by minimality of the lacking position (all other smaller positions must be occupied). 

The total number of moves needed while remaining in each of the planes is bounded by $O(x_My_M)$ and the decrease in $z$ coordinate is instead bounded by $O(z_M)$, giving the desired claim.
\end{proof}

Once in compact form we can reverse this process to reconfigure to our desired configuration $C'$. Hence, with this final lemma, we have proven Theorem~\ref{thm:in-place}. Assuming one of $x_m$ or $y_m$ is odd and the other even, and $n > x_My_M+2z_M $. 

\subsection{Removing Assumptions}\label{sec_details}
In order to complete the proof of Theorem~\ref{thm:in-place} we need to explain how to remove the three assumptions we made at the beginning: the parity of $x_My_M$, how to handle the case in which $n$ is large compared to $x_My_M+z_M$ and how to modify the algorithm so that modules remain in the union of the bounding boxes (rather than the bounding box of the union).

\paragraph*{Fixing Parity}
If the parity of $x_My_M$ is odd, there is no Hamiltonian cycle that we can use to lower the slab one by one. We fix this by virtually ignoring the last x-coordinate. Lower the slab as if it was one row shorter and then lower the remaining row of modules. If $n \geq x_M-1 \times y_M \times z_M$, then both $C$ and $C'$ are dense cubes with fewer empty positions than modules. In this case, it is simple to reconfigure the holes so that they overlap.

Otherwise, our goal is to run the algorithm as if the bounding box were $(x_M-1) \times y_M \times z_M$ with the following changes:

\begin{itemize}
    \item Define the canonical path based on a $(x_M-1) \times y_M$ grid as usual. Note that a module in  $(x_M,y,z)$ cannot be mapped to a point $(\rho_i,z)$.
    \item When building the scaffolding create a slab of size $(x_M-1) \times y_M$.
    \item In the Meld phase we add a small preprocessing step before dropping the slab: any module that is in the line $x=x_M$ and can move does so. They move to any compact position above the slab. If the slab is still at $z_M$ they can move into any position of $B$ outside the $x=x_M$ plane.
    \item We run this preprocessing step each time before dropping the slab. This allows us to obtain following additional invariant: any module at position $(x_m,y,z')$ satisfies $z \leq z_i$.
    \item Once the slab reaches $z_i=0$ we run the preprocessing step one last time before moving to Reconfiguration into Compact form
\end{itemize}

\begin{lemma}
With the modifications mentioned above, the Bounded-Space algorithm will properly run when $x_My_M$ is even.
\end{lemma}
\begin{proof}
First notice that we changed the parity of one of the dimensions without affecting the other, so the Hamiltonian path will now exist. 

The key aspect of correctness is to show that we can indeed remove all modules from the plane $x=x_M$. Indeed, the invariant says that in the worst case a module $m$ in position $(x_m,y,z')$  will be removed when the slab reaches $z_i=z'-1$. Assume, for the sake of contradiction, that this is not true and consider the first case that this condition is violated:

Recall that each time the slab reaches a new coordinate we run a preprocessing step. This applies to when the slab reaches $z_i=z'$. If at that point module $m$ cannot move, it must be a cut vertex and thus there must be a module in position $(x_m,y,z'-1)$ (there cannot be modules above by the invariant, and modules to the side would still be connected to the configuration via the slab). In particular, this implies $z'>1$.

But if this is the case, when the slab drops down to $z_i=z'-1$ we know it can move. This is because the module at $(x_m,y,z'-1)$ previously required $m$ for connectivity, but that module is now adjacent to the slab. In other words, we have created a cycle and thus $m$ will be capable of moving in the preprocessing step executed before dropping the slab to $z_i=z'-2$.  
\end{proof}

\paragraph*{Large Bounding Box Dimensions}
The Scaffolding phase requires $\Omega(x_My_M+z_M)$ many calls to $\LocateAndFree$. Whenever $n< x_My_M+2z_M$ we simply cannot build the scaffolding (since there are not enough modules in the configuration). In this case, we call $\LocateAndFree$ as many times as we can until all modules are part of the scaffolding. When this happens all modules will be in a connected component in the plane $z=z_M$ with 1 or two possibly incomplete vertical lines at $\rho_{i_0}$ and $\rho_{0}$. Note that the initial and target configuration may have a different value for $i_0$ and shapes may even be different. In any case, reconfiguring between the two intermediate shapes is a straightforward modification of the algorithm described in Section~\ref{sec_almostcompact}. 

\begin{lemma}
If $n< x_My_M+2z_M$, the Bounded-Space algorithm runs in $O(n^2)$ time
\end{lemma}

Observe that in this case, the two algorithms are  practically identical as they both run $\LocateAndFree$ $n$ times. The only change is the position in which modules obtained by that procedure are placed.

\paragraph*{Making the Algorithm In-Place}
For simplicity in the description, we considered a single bounding box---i.e., we worked on the bounding box of the union of both configurations rather than considering the union of the two separate bounding boxes, which is the typical framing of in-place. In the following, we make a small variation to fix this discrepancy.

\begin{lemma}
The Bounded-Space algorithm can be modified so that it runs in-place without affecting its runtime. 
\end{lemma}
\begin{proof}
If we put aside the difference in bounding box definition, our algorithm is in-place in the sense that as at most only one module leaves the bounding box, and when it does so it is never more than one unit away.

We reduce the workspace by adding two intermediate configurations: let $B$ be the bounding box of the start configuration $C$ and $B'$ the bounding box of the end configuration $C'$. Let $D$ be the compact configuration of the $C$ with respect to $B$ (respectively $D'$ with respect to $C'$ and $B'$). Note that we can use the in-place algorithm as described between $C$ and $D$ as well as between $D'$ and $C'$. In the process we never move a robot more than a unit away from $B$ or $B'$, respectively. Thus, the algorithm is properly in-place.

Reconfiguring between $D$ and $D'$ is also a straightforward modification of the algorithm described in Section~\ref{sec_almostcompact} (look at the symmetric difference of the two boxes $B$ and $B'$ and move any modules in those spaces). Because modules are in such a compact form, paths we follow can be monotone in the three dimensions, and thus we can bound the number of moves in total by $O(n(x_M+y_M+z_M))= O(nz_M)$.
\end{proof}

\section{Conclusions and Future Work}
The algorithms presented in this paper raise our understanding of the 3D (and higher) sliding cube configuration problem to a level comparable to the 2D counterpart. Although our algorithms are optimal from several points of view (maximum required number of moves, workspace size and input-sensitiveness), several issues remain open:

\begin{itemize}
    \item Due to the practical constraints of physical robots it is desirable for these algorithms to run in a distributed fashion using mostly local information. Although $\LocateAndFree$ transforms connectivity questions into relatively cheap postorder count comparisons, currently this information cannot be easily updated after a module has been freed. Can the update be somehow localized?
    \item The key property for $\LocateAndFree$ is Lemma~\ref{lem:structure}. Since the result is topological, it applies to any dimension $d>2$. Although it would not make much sense to extend the results to higher dimensions, it would be interesting to explore if there are other meaningful module shapes for which the same lemma (possibly with minor variations) applies. 
    \item The current Bounded-Space algorithm compares $n$ to $x_My_M$ and either uses one slab or none (see Appendix). This leads to a serial algorithm, it would be interesting to modify the algorithm to be parallelizable---say, construct $c$ equally spaced slabs (for some $c=c(n,x_My_M,z_M)$), and have each one be responsible for a fraction of the domain. This would not impact the total number of moves but would reduce the \emph{makespan} required to execute them.
    \item As stated in Section~\ref{sec:intro}, sliding cube algorithms can be applied to the crystalline model via $2 \times 2 \times 2$ \textit{meta-modules} (sub-arrangements of modules that are treated as atomic units). Are there meta-modules for other models of modular robots that emulate sliding cubes and can therefore use our algorithms for universal configuration? 
    \item Akitaya et al.~\cite{a.akitaya_et_al:LIPIcs.SWAT.2022.4} proved it was NP-hard to compute the shortest reconfiguration sequence in the 2D sliding square model. Can their result be extended to higher dimensions?
\end{itemize}

\section*{Acknowledgments}
The authors would like to thank Maarten L\"offler and for his contributions during early discussions as well as the authors of~\cite{inPlaceCompact23} and the anonymous reviewers for their valuable comments. Finally, we would like to thank 
Kevin Li and Colton Wolk for implementing preliminary versions of the algorithms proposed in this paper.

Part of this work was conducted during the Simons Laufer Mathematical Sciences Institute Fall 2023 program on the Mathematics and Computer Science of Market and Mechanism Design, which was supported by the National Science Foundation under Grant No.\ DMS-1928930 and by the Alfred P.\ Sloan Foundation under grant G-2021-16778.

\bibliographystyle{plainurl}
\bibliography{Abel-Akitaya-Kominers-Korman-Stock_Sliding_Cube-Shaped_Robots}

\begin{thebibliography}{10}

\bibitem{abel2008universal}
Zachary Abel and Scott~D. Kominers.
\newblock Universal reconfiguration of (hyper-)cubic robots.
\newblock {\em arXiv preprint}, 2008.
\newblock \href {https://arxiv.org/abs/0802.3414v3} {\path{arXiv:0802.3414v3}}.

\bibitem{akitaya2021universal}
Hugo~A. Akitaya, Esther~M. Arkin, Mirela Damian, Erik~D. Demaine, Vida
  Dujmovi{\'c}, Robin Flatland, Matias Korman, Belen Palop, Irene Parada,
  Andr{\'e} van~Renssen Renssen, and Vera Sacristán.
\newblock Universal reconfiguration of facet-connected modular robots by
  pivots: The {$O(1)$ Musketeers}.
\newblock {\em Algorithmica}, 83:1316--1351, 2021.
\newblock \href {https://doi.org/10.1007/s00453-020-00784-6}
  {\path{doi:10.1007/s00453-020-00784-6}}.

\bibitem{a.akitaya_et_al:LIPIcs.SoCG.2021.10}
Hugo~A. Akitaya, Erik~D. Demaine, Andrei Gonczi, Dylan~H. Hendrickson, Adam
  Hesterberg, Matias Korman, Oliver Korten, Jayson Lynch, Irene Parada, and
  Vera Sacrist\'{a}n.
\newblock Characterizing universal reconfigurability of modular pivoting
  robots.
\newblock In Kevin Buchin and \'{E}ric Colin~de Verdi\`{e}re, editors, {\em
  Proceedings of the 37th International Symposium on Computational Geometry
  (SoCG 2021)}, volume 189 of {\em Leibniz International Proceedings in
  Informatics (LIPIcs)}, pages 10:1--10:20, Dagstuhl, Germany, 2021. Schloss
  Dagstuhl -- Leibniz-Zentrum f{\"u}r Informatik.
\newblock \href {https://doi.org/10.4230/LIPIcs.SoCG.2021.10}
  {\path{doi:10.4230/LIPIcs.SoCG.2021.10}}.

\bibitem{a.akitaya_et_al:LIPIcs.SWAT.2022.4}
Hugo~A. Akitaya, Erik~D. Demaine, Matias Korman, Irina Kostitsyna, Irene
  Parada, Willem Sonke, Bettina Speckmann, Ryuhei Uehara, and Jules Wulms.
\newblock Compacting squares: Input-sensitive in-place reconfiguration of
  sliding squares.
\newblock In Artur Czumaj and Qin Xin, editors, {\em Proceedings of the 18th
  Scandinavian Symposium and Workshops on Algorithm Theory (SWAT 2022)}, volume
  227 of {\em Leibniz International Proceedings in Informatics (LIPIcs)}, pages
  4:1--4:19, Dagstuhl, Germany, 2022. Schloss Dagstuhl -- Leibniz-Zentrum
  f{\"u}r Informatik.
\newblock \href {https://doi.org/10.4230/LIPIcs.SWAT.2022.4}
  {\path{doi:10.4230/LIPIcs.SWAT.2022.4}}.

\bibitem{aloupis2009realistic}
Greg Aloupis, S{\'e}bastien Collette, Mirela Damian, Erik~D Demaine, Dania
  El-Khechen, Robin Flatland, Stefan Langerman, Joseph O’Rourke, Val Pinciu,
  Suneeta Ramaswami, Vera Sacrist\'{a}n, and Stefanie Wuhrer.
\newblock Realistic reconfiguration of crystalline (and telecube) robots.
\newblock In {\em Algorithmic Foundation of Robotics VIII: Selected
  Contributions of the Eight International Workshop on the Algorithmic
  Foundations of Robotics}, pages 433--447. Springer, 2009.
\newblock \href {https://doi.org/10.1007/978-3-642-00312-7_27}
  {\path{doi:10.1007/978-3-642-00312-7_27}}.

\bibitem{aloupis2009linear}
Greg Aloupis, S{\'e}bastien Collette, Mirela Damian, Erik~D Demaine, Robin
  Flatland, Stefan Langerman, Joseph O'Rourke, Suneeta Ramaswami, Vera
  Sacrist{\'a}n, and Stefanie Wuhrer.
\newblock Linear reconfiguration of cube-style modular robots.
\newblock {\em Computational Geometry}, 42(6-7):652--663, 2009.
\newblock \href {https://doi.org/10.1016/j.comgeo.2008.11.003}
  {\path{doi:10.1016/j.comgeo.2008.11.003}}.

\bibitem{aloupis2008reconfiguration}
Greg Aloupis, S{\'e}bastien Collette, Erik~D. Demaine, Stefan Langerman, Vera
  Sacrist{\'a}n, and Stefanie Wuhrer.
\newblock Reconfiguration of cube-style modular robots using $o (\log n)$
  parallel moves.
\newblock In {\em Proceedings of Algorithms and Computation: 19th International
  Symposium, ISAAC 2008, Gold Coast, Australia, December 15-17, 2008}, pages
  342--353. Springer, 2008.
\newblock \href {https://doi.org/10.1007/978-3-540-92182-0_32}
  {\path{doi:10.1007/978-3-540-92182-0_32}}.

\bibitem{DP}
Adrian Dumitrescu and J{\'a}nos Pach.
\newblock Pushing squares around.
\newblock {\em Graphs and Combinatorics}, 22(1):37--50, 2006.
\newblock \href {https://doi.org/10.1007/s00373-005-0640-1}
  {\path{doi:10.1007/s00373-005-0640-1}}.

\bibitem{dumitrescu2004motion}
Adrian Dumitrescu, Ichiro Suzuki, and Masafumi Yamashita.
\newblock Motion planning for metamorphic systems: Feasibility, decidability,
  and distributed reconfiguration.
\newblock {\em IEEE Transactions on Robotics and Automation}, 20(3):409--418,
  2004.
\newblock \href {https://doi.org/10.1109/TRA.2004.824936}
  {\path{doi:10.1109/TRA.2004.824936}}.

\bibitem{fitch2003reconfiguration}
Robert Fitch, Zack Butler, and Daniela Rus.
\newblock Reconfiguration planning for heterogeneous self-reconfiguring robots.
\newblock In {\em Proceedings of the 2003 IEEE/RSJ International Conference on
  Intelligent Robots and Systems (IROS 2003)}, volume~3, pages 2460--2467.
  IEEE, 2003.
\newblock \href {https://doi.org/10.1109/IROS.2003.1249239}
  {\path{doi:10.1109/IROS.2003.1249239}}.

\bibitem{inPlaceCompact23}
Irina Kostitsyna, Tim Ophelders, Irene Parada, Tom Peters, Willem Sonke, and
  Bettina Speckmann.
\newblock Optimal in-place compaction of sliding cubes.
\newblock {\em arXiv preprint}.
\newblock \href {https://arxiv.org/abs/2312.15096} {\path{arXiv:2312.15096}}.

\bibitem{miltzow2020hiding}
Tillmann Miltzow, Irene Parada, Willem Sonke, Bettina Speckmann, and Jules
  Wulms.
\newblock Hiding sliding cubes: Why reconfiguring modular robots is not easy
  (media exposition).
\newblock In {\em Proceedings of the 36th International Symposium on
  Computational Geometry (SoCG 2020)}. Schloss Dagstuhl-Leibniz-Zentrum f{\"u}r
  Informatik, 2020.
\newblock \href {https://doi.org/10.4230/LIPIcs.SoCG.2020.78}
  {\path{doi:10.4230/LIPIcs.SoCG.2020.78}}.

\bibitem{moreno2020reconfiguring}
Joel Moreno and Vera Sacrist{\'a}n.
\newblock Reconfiguring sliding squares in-place by flooding.
\newblock In {\em Proceedings of the 36th European Workshop on Computational
  Geometry (EuroCG'20)}, 2020.
\newblock Art.~32.

\bibitem{sung2015reconfiguration}
Cynthia Sung, James Bern, John Romanishin, and Daniela Rus.
\newblock Reconfiguration planning for pivoting cube modular robots.
\newblock In {\em Proceedings of the 2015 IEEE International Conference on
  Robotics and Automation (ICRA)}, pages 1933--1940. IEEE, 2015.
\newblock \href {https://doi.org/10.1109/ICRA.2015.7139451}
  {\path{doi:10.1109/ICRA.2015.7139451}}.

\end{thebibliography}

\end{document}